\documentclass[prb,superscriptaddress,twocolumn,floats,floatfix]{revtex4}
\usepackage{amssymb}
\usepackage{amsmath}
\usepackage{upgreek}
\usepackage{graphicx}
\usepackage{subfigure}
\usepackage[colorlinks=true,citecolor=blue,linkcolor=blue,draft=false]{hyperref}
\usepackage{hyperref}
\usepackage{graphics}
\usepackage{gensymb}
\usepackage{bm}

\begin{document}
\title{Electron spin resonance and spin-valley physics in a silicon double quantum dot}

\author{Xiaojie Hao \footnote{These authors contributed equally to this work}}
\email{haoxj@ucla.edu}
\affiliation{Department of Physics and Astronomy, University of
California at Los Angeles, 405 Hilgard Avenue, Los Angeles,
California 90095, USA}
\author{Rusko Ruskov ${ }^{*}$ }
\email{ruskovr@lps.umd.edu}
\affiliation{Laboratory for Physical Sciences, 8050 Greenmead Dr., College Park, MD 20740, USA}
\author{Ming Xiao \footnote{Present address: Key Laboratory of
    Quantum Information, University of Science and Technology of
    China, Hefei 230026, People's Republic of China}}
\affiliation{Department of Physics and Astronomy, University of
California at Los Angeles, 405 Hilgard Avenue, Los Angeles,
California 90095, USA}
\author{Charles Tahan}
\affiliation{Laboratory for Physical Sciences, 8050 Greenmead Dr., College Park, MD 20740, USA}
\author{HongWen Jiang}
\affiliation{Department of Physics and Astronomy, University of
California at Los Angeles, 405 Hilgard Avenue, Los Angeles,
California 90095, USA}

\date{\today}

\begin{abstract}

  Silicon quantum dots are a leading approach for solid-state quantum
  bits. However, developing this technology is complicated by the
  multi-valley nature of silicon. Here we observe transport of
  individual electrons in a silicon CMOS-based double quantum dot under
  electron spin resonance. An anticrossing of the driven dot
  energy levels is observed when the Zeeman and valley splittings
  coincide. A detected anticrossing splitting of $60\,{\rm MHz}$ is
  interpreted as a direct measure of spin and valley mixing,
  facilitated by spin-orbit interaction in the presence of non-ideal
  interfaces. A lower bound of spin dephasing time of $63\, {\rm ns}$
  is extracted. We also describe a possible experimental evidence of
  an unconventional spin-valley blockade, despite the assumption of
  non-ideal interfaces. This understanding of silicon spin-valley
  physics should enable better control and read-out techniques for the
  spin qubits in an all CMOS silicon approach.

\end{abstract}

\maketitle

\section*{INTRODUCTION}
\label{intro}

It has long been speculated that qubits based on individual electron
spins in Si quantum dots (QDs) have considerable potential for quantum
information processing. Attractive features are the extremely long
coherence time of spins in Si bulk materials and the possibility to
approach zero hyperfine interaction to nuclear spins in
isotopically-purified structures. Furthermore, the extensive
collection of Complementary Metal-Oxide-Semiconductor (CMOS)-based
techniques, accumulated over decades, is
expected to be very important for fabricating many qubits. Electric
and magnetic fields along with charge detection enable qubit gates and
readout. A long coherence time $T_{2}^{*}$ was recently confirmed for
the singlet and triplet states in a Si/SiGe double quantum dot (DQD)
qubit \cite{Maune2012N}.

Electron spin resonance (ESR) is a direct means to drive rotations of
a spin qubit. For electron spins bound in Si, an ensemble of spins in
various structured materials \cite{Morton2011N}, single electrons in a
single defect \cite{Xiao2004N}, and in a single donor \cite{Pla2012N}
have been explored with ESR, using various detection schemes. Physical
implementations of ESR on individual bound electronic spins have
proven to be successful in GaAs-based QDs transport experiments
\cite{Koppens2006N,Pioro-Ladriere2008NP}, where the essential role of
the spin (Pauli) blockade and the nuclear spin bath in that systems
were established. However, spin detection via electronic transport in
gate defined Si QDs has remained challenging.

Here, we report the detection of microwave driven electron spin
resonance transport of individual electrons in a silicon
metal-oxide-semiconductor (MOS) based DQD. The lifting of the blockade
via ESR is detectable only in a narrow region where Zeeman split spin
states of different valley content anticross where the Zeeman
splitting $E_Z$ equals the valley splitting of $E_V \simeq 86.2\,
\mu{\rm eV}$. We show that the anticrossing is due to spin-orbit
coupling in the heterostructure, in the presence of interface
roughness, that mixes spin and valley states (similar mixing mechanism
was first established in the electron spin relaxation in a small
single Si quantum dot\cite{Yang2013NC}). The gap at anticrossing of
$\Delta f_{\rm anti-cros} \simeq 60\,{\rm MHz}$ is a measure of this
spin-valley mixing and also provides a means to access higher valley
states via ESR. Analysis of the ESR spectrum provides a lower-bound
estimation of an inhomogeneous decoherence time, $T_{2}^{*} \simeq
63\,{\rm ns}$, which is much longer than that in GaAs
\cite{Petta2005S}; it also compares well to the direct measurement of
$T_{2}^{*}$ for a Si qubit encoded by singlet-triplet states
\cite{Maune2012N}. The nature of the experimental blockade regime, and
its dependence on the applied magnetic field and interdot energy
detuning is discussed. Since the observed blockade region $0 <
\varepsilon < \Delta_{\rm ST}^{\rm exp} \simeq 343\, \mu{\rm eV}$
includes detunings larger than the valley splitting we can conclude
that a spin-valley blockade takes place, related to the
impossibility for an inelastic (via phonons) electron tunneling to
happen. This blockade survives even in the presence of a non-ideal
interface. The observations made in this paper encourage further
development of Si MOS-based spin qubits and further suggest the
additional valley degree of freedom
\cite{Maune2012N,Goswami2007NP,Culcer2012PRL,Culcer2010PRB2,Yang2013NC,Zhang2013NC}
is critical to understanding silicon qubits.

\section*{RESULTS}

\subsection*{DQD device}

\begin{figure}[!tbp]
    \centering{\includegraphics[]{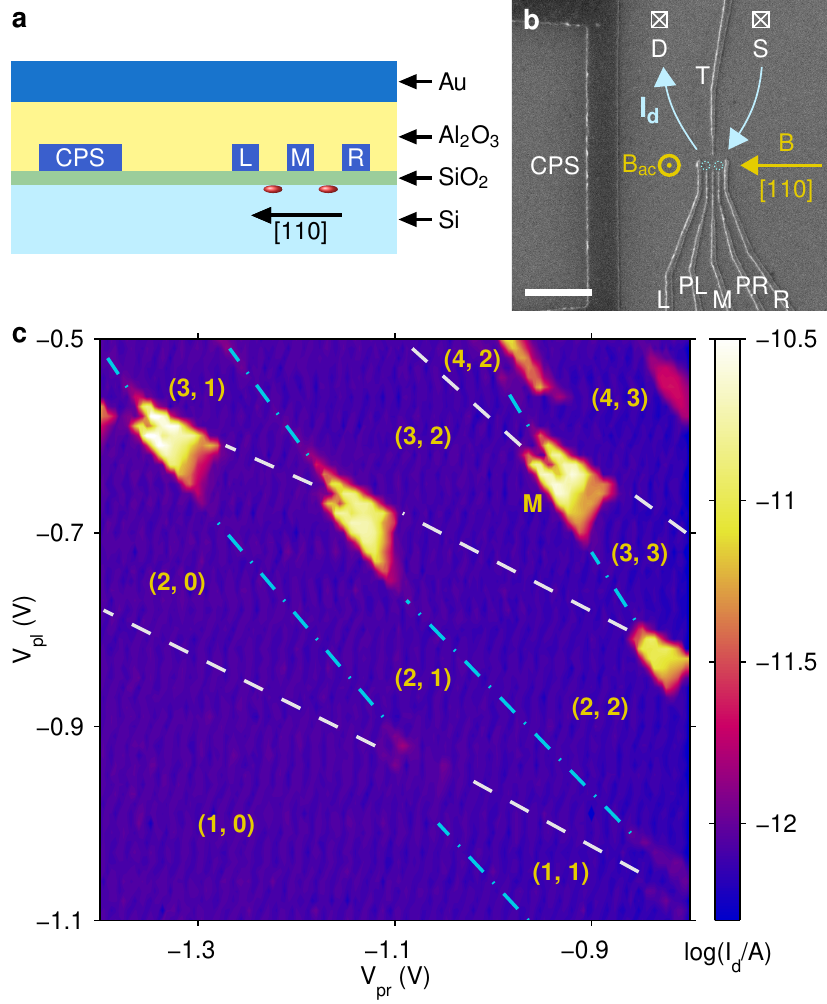}}
    \caption{\textbf{Silicon MOS double quantum dot.} (\textbf{a}) The
      cross-sectional view of the device. (\textbf{b}) The SEM image
      of a similar device (scale bar: $1\,\mu{\rm m}$.) Double quantum
      dot (blue circles) is defined at the $\mathrm{Si/SiO}_{2}$
      interface by the confinement gates (labeled as `T', `L', `PL',
      `M', `PR', and `R'). Microwave applied to CPS loop generates an
      AC magnetic field at the quantum dot. (\textbf{c}) Stability
      diagram at few electron region (log scale), with additional
      labels for estimated electron numbers and guidelines for
      transitions between different electron configurations.}
\label{fig:1}
\end{figure}

The cross-sectional view of the Si MOS QD device is shown in
Fig.~\ref{fig:1}a. A scanning electron microscope (SEM) image of the
essential part of a similar device is shown in Fig.~\ref{fig:1}b. A
double quantum dot (DQD) is defined by six confinement gates, which
are labeled as `T', `L', `PL', `M', `PR', and `R'. A coplanar strip
(CPS) loop (see Methods for device details), situated about $1.5$
microns away, is used to deliver an oscillating (AC) magnetic field
$B_{\rm ac}$, perpendicular to the DQD interface. The oscillation
frequency $f_{\rm ac}$ is scanned to resonance with the electron spin
precession oscillations in an in-plane external magnetic field $B$,
Fig.~\ref{fig:1}b. The DQD is characterized by the DC transport
current. Fig.~\ref{fig:1}c shows a typical charge stability diagram of
the device with source-drain bias voltage of $V_{\rm sd} = -1\,{\rm
  mV}$, in which the transport current is recorded while the plunger
gates $V_{pl}$ and $V_{pr}$ are scanned \cite{Wiel2002RMP}. The device
does not contain a charge sensing channel and the identified electron
numbers are the approximate one. The estimated electron occupation
numbers in the left and right dots are labeled by $(N_{L}, N_{R})$. At
lower electron numbers (more negative voltage at the plunger gates
`PL', `PR') the tunneling from (out of) source (drain) is suppressed,
so higher biasing triangles will be examined on electronic transport.
Electron transitions into and out of the left (right) dot are labeled
by white dashed (blue dash-doted) lines in Fig.~\ref{fig:1}c. The
honeycomb structure and the biasing triangles here show the
characteristic features of a well defined DQD \cite{Wiel2002RMP}.

\subsection*{Spin blockade}

\begin{figure}[!tbp]
    \centering{\includegraphics[]{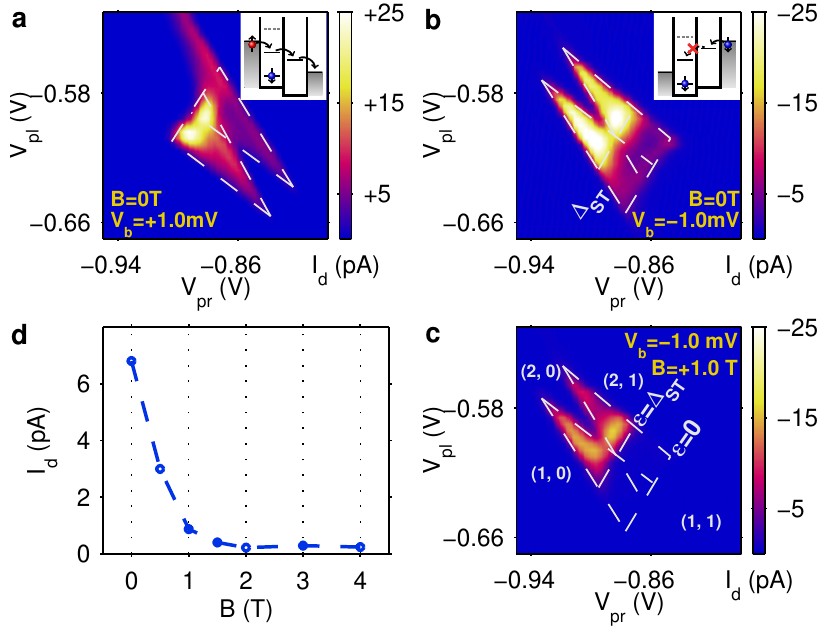}}
    \caption{\textbf{Spin blockade at biasing triangle labeled as `M'
        in Fig.~\ref{fig:1}c.} (\textbf{a}) and (\textbf{b}),
      Transport data at forward and reverse bias voltage applied to
      source and drain (external magnetic field $B=0\,{\rm T}$). The
      low current region in (\textbf{b}) is a signature of spin
      blockade. Insets demonstrate that at forward bias voltage, only
      spin up can tunnel into the left dot and then tunnel out, which
      contributes to the transport current, while at reverse bias
      voltage, the transport is prohibited by Pauli exclusion
      principle once a spin down electron tunnels into the right dot.
      (\textbf{c}) Transport data at reverse bias voltage with
      external magnetic field on ($B=1\,{\rm T}$). Valence electron
      numbers are labeled. The leakage current in the spin blockade
      region is well suppressed by the magnetic field. (\textbf{d})
      Leakage current in spin blockade region measured at different
      magnetic fields. }
\label{fig:2}
\end{figure}

Spin blockade of the electronic transport is the well known method for
sensing and manipulation of confined electron spins in semiconductor
heterostructures \cite{Ono2002S,Petta2005S,Koppens2006N,Maune2012N}.
For a DQD confining two electrons, the standard statement is that an
electron cannot flip spin under tunneling, and so a transition from a
$(1,1)$ charge configuration to a $(2,0)$ configuration is only
possible between the corresponding singlet or triplet spin states:
$\rm S{(1,1)} \to S{(2,0)}$, $\rm T{(1,1)} \to T{(2,0)}$, respecting
the Pauli exclusion principle (see the Fig.~\ref{fig:2}a,b, insets).
In a typically biased DQD (with detuning much larger than tunneling,
$\varepsilon \gg t_c$), the delocalized states $\rm S{(1,1)}$, $\rm
T{(1,1)}$ are only slightly shifted by an exchange energy $J \simeq
2t_c^2/\varepsilon$, while the localized states $\rm S{(2,0)}$, $\rm
T{(2,0)}$ are split by large $\Delta_{ST} \gg J$ (given a higher
orbital excitation of a few hundred $\mu{\rm eV}$). Thus, the energy
of the state $\rm T{(2,0)}$ is much higher, and transition of $\rm
T{(1,1)}$ to $\rm T{(2,0)}$ is energetically forbidden, while
transition of $\rm S{(1,1)}$ to $\rm S{(2,0)}$ is still allowed.

In Si the conduction electrons may belong to different valley
configurations $\rm v_i$, ($i=1,2$ at the 2D interface): the electron
state acquires a valley index, $\rm v_1$ or $\rm v_2$, denoting two
states that have the same spin and orbital content, but are split
by a valley energy $E_V$; we further refer to these states as valley
states, though they are certain superposition of the original valleys
\cite{Tahan2002PRB66}, $\pm\hat{z}$, (see Supplementary Note 1). Thus,
the simple spin-blockade can be applied to the lowest $\rm
[v1,v1]$-valley states. In what follows, we first make an attempt to
describe the experimental blockade and ESR following the simple
spin-blockade picture. The implications of higher valley states, $\rm
[v1,v2]$, $\rm [v2,v2]$, will be considered later (also see
Supplementary Notes 2, 3 and 4).

We focus on the biasing triangle labeled as `M' in Fig.~\ref{fig:1}c,
which occurs at the transition between the electron states $(4,2)$
and $(3,3)$. Assuming only ``valence'' electron configurations take
place in the transport \cite{Shaji2008NP,Lai2011SR}, we use hereafter
the effective electron occupancy $(2,0)$ and $(1,1)$, as labeled in
Fig.~\ref{fig:2}c. As a key signature of the spin blockade, the
forward-bias (Fig.~\ref{fig:2}a) transport is allowed within the whole
detuning region, while the reverse-bias (Fig.~\ref{fig:2}b) transport
shows a low current region\cite{Johnson2005PRB}. As illustrated, at
forward bias (inset of Fig.~\ref{fig:2}a), only spin singlet $\rm S(2,
0)$ can be formed when the second electron tunnels into the left dot
(DQD is in the $(1,0)$ configuration before tunneling), and then the
$\rm S(2,0)$ state can make transition to $(1,0)$ state through a
$\rm S(1,1)$ state. Therefore, a continuous flowing transport current
will be observed. However, at reverse bias (inset of
Fig.~\ref{fig:2}b), once a triplet state $\rm T(1,1)$ is formed, it
cannot make transition to a $(2,0)$ charge state, and thus blocks the
current \cite{Ono2002S,Johnson2005PRB}. With a finite magnetic field
applied (specifically, in this experiment we used an in plane field,
parallel to the DQD, and oriented along the $[110]$
crystallographic direction at the Si interface, Fig.~\ref{fig:1}a and
Fig.~\ref{fig:1}b), the triplet $\rm T(1,1)$ state splits into three
states: $\rm T_{+}(1,1)$, $\rm T_{0}(1,1)$ and $\rm T_{-}(1,1)$,
and the current can be blocked by loading electrons into any of these
three states. The low current (blocked) region, traced by the white
trapezoid in Fig.~\ref{fig:2}b, implies a $(2,0)$ singlet-triplet
splitting of $\Delta^{\rm exp}_{\rm ST} = 343 \pm 29\,\mu{\rm eV}$
(see, however, the discussion of spin-valley blockade below).

The current in the spin blockade region in Fig.~\ref{fig:2}b is not
completely suppressed; in real systems, the electron spins can be
mixed or flipped by the nuclear field hyperfine interaction
\cite{Koppens2005S,Nadj-Perge2010PRB}, spin-orbit coupling
\cite{Nadj-Perge2010PRB,Yamahata2012PRB} or co-tunneling
\cite{Lai2011SR,Coish2011PRB,Yamahata2012PRB,Koh2011PRL}, which
generates a finite leakage current and lifts the spin blockade. This
leakage can be strongly suppressed, in Fig.~\ref{fig:2}c, by applying
an external magnetic field \cite{Coish2011PRB,Lai2011SR}, which is
parallel to the $\mathrm{Si/SiO_2}$ interface. Almost one order of
magnitude suppression of the leakage current, as shown in
Fig.~\ref{fig:2}d, allows us to probe the electron spin at higher
field ($B \gtrsim 0.5\,{\rm T}$) with a good sensitivity.

\begin{figure}[tbp]
    \centering{\includegraphics[]{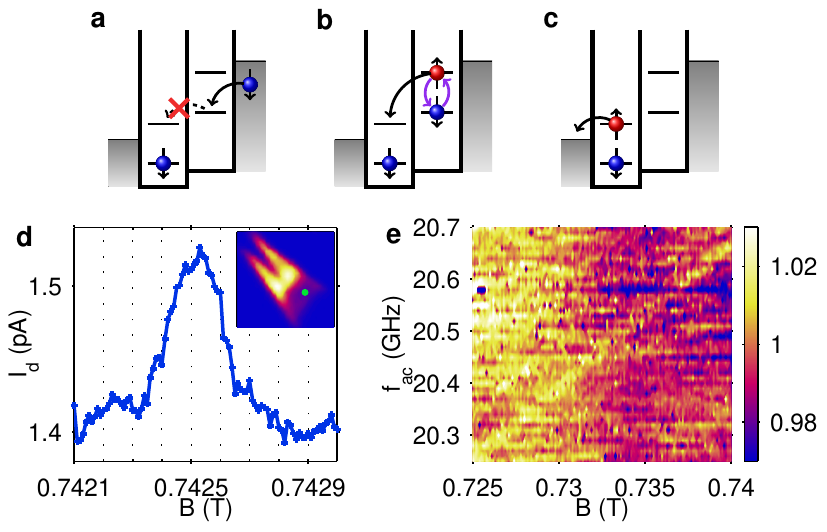}}
    \caption{\textbf{Single electron spin resonance scheme and data.}
      (\textbf{a})-(\textbf{c}) Single electron spin resonance
      detection scheme. (\textbf{a}) An additional electron tunnels
      into the right dot but transport is blocked due to Pauli
      exclusion principle. (\textbf{b}) Microwave drives electron spin
      in resonance hence opens a transport channel. (\textbf{c})
      Second electron with flipped spin tunnels out, contributing to
      the transport current. (\textbf{d}) A typical ESR signal
      obtained by fixing microwave frequency at $f_{ac} = 20.77\,{\rm
        GHz}$ (microwave power $P= -28\,{\rm dBm}$) while scanning
      magnetic field (averaged over 20 curves). Inset: green dot in
      the biasing triangle (at $B = 0.7425\,{\rm T}$, frequency
      $f_{ac} = 20.71\,{\rm GHz}$, and microwave power $P= -20\,{\rm
        dBm}$) shows where the ESR is detected. (\textbf{e})
      Normalized leakage current (see text) as a function of magnetic
      field and microwave frequency. ESR peak position shows linear
      relation between magnetic field and microwave frequency. }
\label{fig:3}
\end{figure}

\subsection*{Detection of ESR and phase coherence time}
\label{Sec:ESR_detection_coherence_time}

The single electron spin resonance is observed by setting the DQD in
the spin blockade region ( Fig.~\ref{fig:3}a, green dot in the inset
of Fig.~\ref{fig:3}d, corresponding to an interdot detuning
$\varepsilon \approx 100\, \mu{\rm eV}$, in this case), and applying
an oscillating magnetic field via the CPS loop \cite{Koppens2006N}. At
a frequency where the microwave energy matches the Zeeman splitting of
a single electron spin ($hf_{ac}=g\mu_{B}B$, where $\mu_B$ is Bohr
magneton and $g$ is the electron $g$-factor), the spins can flip.
Since, however, the AC field rotates the spins in the two dots
simultaneously, the two electrons will remain within the triplet
subspace and the spin blockade will not be lifted (see also the
Supplementary Note 3). Thus, a spin mixing mechanism is required to
mix the triplet $(1,1)$ subspace with the singlet $(1,1)$
(Fig.~\ref{fig:3}b), that can subsequently inelastically tunnel to a
$(2,0)$ state, lifting the blockade (Fig.~\ref{fig:3}c). Similar to a
GaAs DQD system \cite{Koppens2006N}, an inhomogeneous nuclear
hyperfine (HF) field $\sigma_N$ could mix the singlet $\rm S(1,1)$ and
triplets $\rm T(1,1)$, assuming the HF energy, $E_N \equiv g\mu_B
\sigma_N$, is larger than the singlet-triplet exchange splitting, $E_N
> J \simeq 2 t_c^2/|\varepsilon|$ (see, e.g.,
Refs.~\onlinecite{JouravlevNazarov2006PRL,Taylor2007PRB76}). While, at
a finite external magnetic field $B$, the polarized triplets $\rm
T_{+}(1,1)$, $\rm T_{-}(1,1)$ are spin blocked since their HF mixing
with the singlet is energy suppressed, an AC field resonant to the
Zeeman energy splitting brings them in resonance with the $\rm
T_{0}(1,1)$ state that can mix to the singlet $\rm S(1,1)$. In a
Si-based system, however, the inhomogeneous nuclear field is one to
two orders of magnitude smaller than that in GaAs system, and one
would expect a much weaker ESR signal.

The ESR peak is observed within a relatively narrow
microwave frequency range, $\Delta f_{\rm ac} \le 1\,{\rm GHz}$, by
measuring the leakage current in the spin blockade region as a
function of the external magnetic field. On Fig.~\ref{fig:3}d, it is
shown for a fixed microwave frequency ($f_{\rm ac} = 20.77\,{\rm
  GHz}$). We have verified that the ESR signal can only be detected in
the spin blockade region of the reverse biasing triangle in
Fig.~\ref{fig:2}b, and persists up to $500\,{\rm mK}$. By measuring
the leakage current as a function of the magnetic field and microwave
frequency, the ESR signal is presented as a sloped straight line in
the two-dimensional space (Fig.~\ref{fig:3}e). For better contrast,
the leakage current in Fig.~\ref{fig:3}e is normalized by the average
current at each frequency. The slope of the ESR line gives an
effective g-factor of $g=1.97\pm 0.07$, compatible to Si. The
relatively large error is since the ESR signal is only visible in a
narrow frequency window. Surprisingly, the ESR signal is just as
strong as that in a GaAs system \cite{Koppens2006N}, where there is a
much larger nuclear HF field.

\begin{figure}[tbp]
    \centering{\includegraphics[]{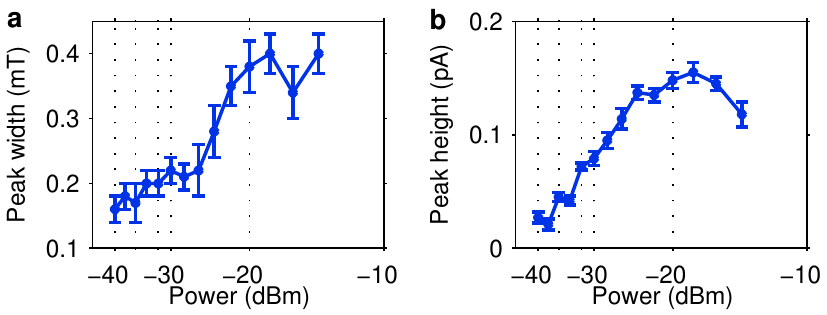}}
    \caption{\textbf{Power dependence of ESR peak.} ESR signal
      (\textbf{a}) line width and (\textbf{b}) peak height (with error
      bars) as a function of applied microwave power. Both of them
      show power saturation. The microwave power is measured at the
      output of the microwave generator. }
\label{fig:4}
\end{figure}

The line width ($\Delta B_{\rm ESR} \equiv$ FWHM) of the ESR peak in
Fig.~\ref{fig:3}d is $\Delta B_{\rm ESR} = 0.2\,{\rm mT}$, which is
one order of magnitude smaller compared to the HF field value of $\sim
2\,{\rm mT}$ measured in GaAs QD\cite{Koppens2006N}. The power
dependence of the ESR line width is plotted in Fig.~\ref{fig:4}a,
showing weak dependence at low power, and power broadening by the
applied AC magnetic field, and finally a saturation
\cite{Koppens2007JAP,Poole1996}. The saturation effect is also seen in
the power dependence of the ESR peak height as shown in
Fig.~\ref{fig:4}b: the ESR peak firstly increases linearly as the
power increasing \cite{Koppens2006N,Engel2001PRL}, and finally
decreases at higher power due to the disturbance of additional
electric field or photon-assisted tunneling.

At low microwave power, the line width of the ESR peak would be
determined by the nuclear field fluctuation, assuming that this
mechanism dominates. The narrowest ESR line below the saturation we
observed is around $\Delta B_{\rm ESR} = 0.15\,{\rm mT}$, giving an
estimated nuclear HF field (in $z$-direction) in a single dot
$\sigma_{N,z} = \Delta B_{\rm ESR}/2\sqrt{\ln{2}} \simeq 0.090\,{\rm
  mT} $. This value gives a lower bound for the inhomogeneously
broadened spin dephasing time (assuming a singlet-triplet qubit)
$T_{2,ST}^{*}=\hbar 2\sqrt{\ln{2}}/(g\mu_{B}\Delta B_{\rm ESR})\approx
63\,{\rm ns}$ ($\hbar$ is the reduced Planck constant), which is
significantly longer than that measured in GaAs DQD system
($T_{2,ST}^{*,{\rm GaAs}} \simeq 10\,{\rm ns}$) via free induction
decay \cite{Petta2005S}.

The value of the estimated hyperfine field, however, is $6$ times
larger than that expected \cite{Assali2011PRB} in Si (with natural
abundance of $\mathrm{ {}^{29}Si}$ nuclei), and also confirmed
experimentally for SiGe quantum dots \cite{Maune2012N}. The
discrepancy may be explained by noting that in our Si DQD with
transport measurement setup, the inelastic interdot tunneling rate may
happen to be large: $\Gamma_{\rm in} \gtrsim g\mu_{B}
\sigma_{N,z}/\hbar \sim 10^7\, {\rm s}^{-1}$, so that the ESR line may
be broadened by the inelastic transition itself (see Supplementary
Note 5). Assuming only this broadening mechanism, the experimental
FWHM-ESR, $\Delta B_{\rm ESR}$ allows to estimate $\Gamma_{\rm in}
\approx 2.6\times 10^7\, {\rm s}^{-1}$, comparable to recent
measurements in $\mathrm{Si/SiO_2}$ DQDs \cite{Wang2013PRL}.

\subsection*{Anticrossing feature in the ESR spectrum}
\label{Sec:anticrossing_feature}

\begin{figure}[tbp]
    \centering{\includegraphics[]{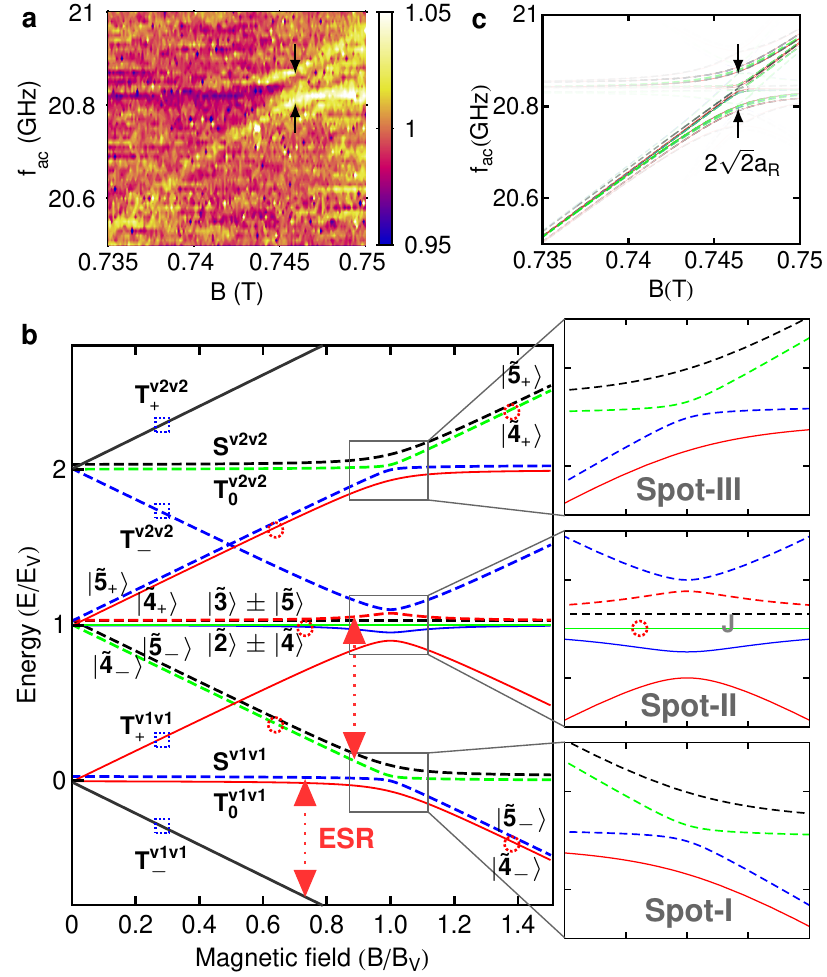}}
    \caption{\textbf{ESR spectroscopy in a Si MOS DQD.} (\textbf{a})
      Normalized leakage current as a function of the external
      magnetic field and microwave frequency. ESR lines show
      anticrossing feature with a frequency gap $\Delta f_{\rm
        anti-cross} =60\pm 10\, {\rm MHz}$ (black arrows), at a
      position $B_V \simeq 0.746\,{\rm T}$ (Methods). (\textbf{b})
      Left panel: field dependent energy diagram of electrons in the
      $(1,1)$ charge states in case of small exchange energy: $J <
      |a_R|$ (see text). Dotted (blue) rectangles label the
      spin-blocked states that are similar to GaAs, but with valley
      index added; Dotted (red) circles label the spin-valley blocked
      states: $\rm |\tilde{4}\rangle = T_0^{\rm v1v2} + T_0^{\rm
        v2v1}$, $\rm |\tilde{4}_{\pm}\rangle = T_{\pm}^{\rm v1v2} +
      T_{\pm}^{\rm v2v1}$, eventually surviving for a non-ideal
      interface (see text). The other (non-blocked) states are: $\rm
      |\tilde{5}\rangle = -T_0^{\rm v1v2} + T_0^{\rm v2v1}$, $\rm
      |\tilde{5}_{\pm}\rangle = -T_{\pm}^{\rm v1v2} + T_{\pm}^{\rm
        v2v1}$, and $\rm |\tilde{2}\rangle = S^{\rm v1v2} - S^{\rm
        v2v1}$, $\rm |\tilde{3}\rangle = S^{\rm v1v2} + S^{\rm v2v1}$.
      Right panels: zoom in of the level anticrossing of states with
      different valley content due to SOC; the size of the energy gaps
      is given by only two dipole matrix elements, $a_R$, $b_L$ (see
      text). (\textbf{c}) Simulated anticrossing features in the ESR
      spectroscopy, using the measured valley energy splitting $E_V =
      86.2\, \mu{\rm eV}$, and the anticrossing splitting of
      Eq.~(\ref{splitting}) (second splitting is invisible for $|b_L|
      \ll |a_R| $). }
    \label{fig:5}
\end{figure}

By exploring the two-dimensional ESR spectroscopy (see
Fig.~\ref{fig:3}e) at higher magnetic field and microwave frequencies,
we notice that the straight ESR line splits and forms an anticrossing
feature at a frequency $f_{\rm anti-cross} \simeq 20.84\,{\rm GHz}$;
also notice a ``remnant'' of the straight ESR line in between the
anticrossing (Fig.~\ref{fig:5}a). The anticrossing position
(corresponding to an energy difference of $h f_{\rm anti-cross}\simeq
86.2\, \mu{\rm eV}$) is determined to be independent of the interdot
detuning $\varepsilon$ (e.g., in the range of $\varepsilon = 50 -
250\, \mu{\rm eV}$ the anticrossing frequency does not shift at
different detunings within an error bar of $20\,{\rm MHz}$). The size
of the anticrossing gap can be readily obtained from the f-B diagram
to be about $\Delta f_{\rm anti-cross} \simeq 60 \pm 10\,{\rm MHz}$.
Similar ESR anticrossing features were observed for a different
biasing triangle than that mentioned on Fig.~\ref{fig:1}c, (see
Supplementary Note 3). Since the experiment involves two coupled
quantum dots, at first sight, one would relate the anticrossing
feature with the level crossing of, e.g., the $\rm T_{-}(1,1)$ and
$\rm S(2,0)$ states at finite magnetic field, if a spin-orbit
interaction would dominate the anticrossing \cite{Danon2009PRB}.
However, the independence on detuning rules out this possibility, as
such crossing would be a strong function of detuning
\cite{Nadj-Perge2012PRL}. Therefore, to explain the anticrossing one
should only include the states with the same charge configuration.

\subsection*{Interpretation of the anticrossing feature}
\label{Sec:interpretation_anticrossing}

Because the independency on detuning of the anticrossing, we will
identify it as due to anticrossing of singlet and triplet $(1,1)$
states of {\it different} valley content. The 2-electron states now
acquire an additional valley index $\rm [v_i,v_j],\, i,j=1,2$. For
zero magnetic field one has three groups of degenerate levels denoted
as $\rm [v1,v1]$, $\rm [v1,v2]$, $\rm [v2,v2]$ (16 states in
total\cite{Culcer2010PRB}, see Supplementary Note 1), that are split
off each other by the valley splitting energy of $E_V = h f_{\rm
  anti-cross}\simeq 86.2\, \mu{\rm eV}$ measured in the experiment (we
have neglected for the moment the finite exchange splitting $J$, as
well as the small nuclear HF field in Si, see below). The polarized
triplet states from each group (anti)cross the singlet(s)/triplet(s)
of the other group at a magnetic field $B_V = E_V/(g\mu_B)\simeq
0.746\,{\rm T}$ (see Supplementary Note 2). For example, the triplet
state $\rm T_{+}^{\rm v1v1}$ will anticross the upper singlets $\rm
S^{\rm v1v2}$, $\rm S^{\rm v2v1}$, and the unpolarized triplets $\rm
T_0^{\rm v1v2}$, $\rm T_0^{\rm v2v1}$ (see Fig.~\ref{fig:5}b, and
Supplementary Figures 1, 2, 3 for other situations). Simultaneously,
for $B \sim B_V$, the triplets $\rm T_{-}^{\rm v1v2}$, $\rm T_{-}^{\rm
  v2v1}$ anticross the lower levels $\rm S^{\rm v1v1}$, $\rm T_0^{\rm
  v1v1}$, and the triplets $\rm T_{+}^{\rm v1v2}$, $\rm T_{+}^{\rm
  v2v1}$ anticross the upper levels $\rm S^{\rm v2v2}$, $\rm T_0^{\rm
  v2v2}$. Thus, three anticrossing ``spots'' are formed at a Zeeman
splitting $E_Z = E_V$, Fig.~\ref{fig:5}b. Since the source-drain
voltage is large, $e V_{\rm sd} \simeq 1\,{\rm meV} \gg E_V$, all
$(1,1)$ states are loaded, and the ESR resonance transport takes place
for the different groups of transitions.

The mixing mechanism of different spin-valley states is due to the
spin-orbit coupling (SOC) in the presence of non-ideal
$\mathrm{Si/SiO_2}$ interface \cite{Yang2013NC}; here it can be
parameterized by only two dipole matrix elements
\begin{eqnarray}
&& a_R = \frac{m_t E_V^R (\beta_D - \alpha_R)}{2\hbar}\, \langle \tilde{R}_{\rm v_1}({\bold r})|(x + y)| \tilde{R}_{\rm v_2}({\bold r}) \rangle
\qquad
\label{dipoleMEa}\\
&& b_L = \frac{m_t E_V^L (\beta_D - \alpha_R)}{2\hbar}\, \langle \tilde{L}_{\rm v_1}({\bold r})|(x + y)| \tilde{L}_{\rm v_2}({\bold r}) \rangle
\qquad
\label{dipoleMEb}
\end{eqnarray}
where $\beta_D$ ($\alpha_R$) is the Dresselhaus (Rashba) SOC parameter
(also, valley splitting may be different in the right/left dot), $m_t
= 0.198 m_e$ is the transverse effective mass for conduction
electrons, and, e.g., $\tilde{R}_{\rm v_1}({\bold r})$ is the wave
function of an electron confined in the right dot
\cite{Burkard1999PRB}. We note that for an ideally flat
interface the above matrix elements (m.e.) are exactly zero. However,
in the presence of disorder/roughness the valley envelop functions of
both valleys are perturbed \cite{Yang2013NC}, providing non-zero
dipole matrix elements; the latter just parameterize the presence
  of a kind of roughness, i.e. a non-flat interface, atomic steps, or
  defects. The splitting at anticrossing is directly observable
via ESR, and (for $|a_R| \gg |b_L|$) is given by
\begin{equation}
\Delta_{\rm anti-cross} \simeq  2\sqrt{2} |a_R|      \label{splitting} ,
\end{equation}
providing a dipole m.e., $x_{12}\equiv \langle \tilde{R}_{\rm
  v_1}({\bold r})| x | \tilde{R}_{\rm v_2}({\bold r}) \rangle$, of the
order of $x_{12} \sim 15-55\, {\rm nm}$ for the measured gap (see
Supplementary Note 2). We note that the same matrix elements were
recently shown to be responsible for spin-valley mixing and fast
(phonon) relaxation of spin states at the so-called ``hot spot'' in a
single $\mathrm{Si/SiO_2}$ QD \cite{Yang2013NC} (see also recent
calculations of such m.e. that confirm their order of magnitude value,
by modeling the interface roughness with single atomic steps at the QD
interface \cite{Gamble2013PRB}).

The description of the ESR and the spin blockade, in
general, are more involved since $(1,1)$ states of various valley
content are loaded, starting from a $(1,0)$ state. For a fixed
external magnetic field there are several transitions in resonance
with the AC oscillating magnetic field, Fig.~\ref{fig:5}b. For the
direction of the magnetic field chosen in the experiment
(Fig.~\ref{fig:1}b) the AC Hamiltonian is: $\hat{H}_{\rm ac} = \frac{g
  \mu_B B_{\rm ac}}{\sqrt{2}} \cos{\omega t} \left[ -i |{\rm
    T_{+}}\rangle \langle {\rm T_0} | + i |{\rm T_{-}}\rangle \langle
  {\rm T_0} | + {\rm h.c.} \right] \otimes {\rm I}^{\rm v_i v_j}$, and
couples triplet states within each valley subspace (both the AC and
nuclear HF interaction do not mix different valley subspaces).
Therefore, the AC coupling between each pair of resonant eigenstates
will be dependent on the projection of these states to the
corresponding triplets, $\rm T_{\pm}$, $\rm T_{0}$, making some of the
transitions suppressed.

The contribution to the ESR signal of the valley subspaces $\rm
[v1,v1]$, $\rm [v2,v2]$ (Fig.~\ref{fig:5}b) is similar to that in a
GaAs DQD \cite{Petta2005S,Koppens2006N} while in a regime where
nuclear HF mixing is suppressed. Even in the absence of (random)
nuclear HF field mixing of $\rm S$-$\rm T_0$ states, the spin-valley
mixing mechanism allows a finite ESR signal to be observed. E.g., for
the spin-blocked state $\rm T_{-}^{\rm v1v1}$, a resonant ESR
transition to the upper state is possible, since it is a coherent
mixture of the states $\rm T_{0}^{\rm v1v1}$ and $\rm S^{\rm v1v1}$ at
the anticrossing spot I (Fig.~\ref{fig:5}b). Thus, the ESR leakage
current will increase at anticrossing due to relative increase of a
singlet (or triplet, see below) state, that can tunnel inelastically
to a $(2,0)$ state. Since the size of the splitting of $60\,{\rm MHz}$
is equivalent to an ``effective magnetic field'' of $\sim 2.1\,{\rm
  mT}$, this explains qualitatively that the observed ESR signal is as
strong as for GaAs DQDs, even though the nuclear HF field in Si is two
orders of magnitude smaller than in GaAs. Out of anticrossing the ESR
signal decreases as well as the $\rm S$-$\rm T_0$ mixing; numerically,
e.g., for the lowest hybridized state $\rm \widetilde{T}_{0}^{\rm
  v1v1}$ (Fig.~\ref{fig:5}b) a ten times smaller admixing of
non-blocked states ($\rm S^{\rm v1v1}$, $|\tilde{5}_{-}\rangle$ in
this case) corresponds to Zeeman detuning $|E_Z - E_V| \approx 7
\Delta_{\rm anti-cross}$, explaining a ``bright ESR range'' of $\sim 2
|E_Z - E_V| \approx 0.8\, {\rm GHz}$ (or $\sim 0.03\, {\rm mT}$),
comparable to the experimentally observed range, Fig.~\ref{fig:5}a.

Another mechanism of ESR signal suppression (especially of the sloped
ESR line) is due to the finite exchange energy splitting. A finite
exchange splitting, $J\simeq 2 t_c^2/|\varepsilon|$, lifts the
singlet-triplet degeneracy for each group of valley states, far from
the anticrossing region, and forms eigenstates (Fig.~\ref{fig:5}b),
where some of them will be blocked. For a coherent tunneling of $t_c
\approx 5-10\, \mu{\rm eV}$ the estimated exchange splitting is in the
range of $J \approx 0.2-0.8\, \mu{\rm eV}$, so it is much larger than
the nuclear HF energy in Si, $E_N \approx 3\,{\rm neV}$. (Even though
we do not measure $J$ directly, a situation when $J < E_N$ is
unlikely, since in this case the ESR suppression out of anticrossing
cannot be explained, see Supplementary Note 3). Thus, the standard
mechanism of a $\rm S$-$\rm T_0$ mixing via the HF field will be
energetically suppressed far from anticrossing for the $\rm [v1,v1]$
and $\rm [v2,v2]$ valley states, and so its corresponding contribution
to the observed ESR signal.

Since in the ESR transport experiment upper valley states are loaded,
one need to consider one more mechanism of ESR signal suppression. We
assume that within the $\rm [v1,v2]$ valley subspace, the polarized
state $\rm |\tilde{4}_{-} \rangle = T^{\rm v1v2}_{-} + T^{\rm
  v2v1}_{-}$ is spin-valley blocked (see the discussion of
spin-valley blockade below). Since out of anticrossing it is
equally coupled via $\hat{H}_{\rm ac}$ to two degenerate states
$\rm |\tilde{2}\rangle \pm |\tilde{4}\rangle$, Fig.~\ref{fig:5}b, this
creates a coherent superposition of these states, in which the singlet
part, $\rm |\tilde{2}\rangle = S^{\rm v1v2} - S^{\rm v2v1}$, is
canceled, while the triplet $\rm |\tilde{4} \rangle = T^{\rm v1v2}_{0}
+ T^{\rm v2v1}_{0}$ is spin-valley blocked. Further HF coupling of
$\rm |\tilde{4} \rangle$ to the unblocked singlet $\rm |\tilde{3}
\rangle = S^{\rm v1v2} + S^{\rm v2v1}$, is suppressed by the finite
exchange splitting $J$ (Fig.~\ref{fig:5}b, left and right panels), and
so is the ESR signal. The above arguments complete the explanation of
suppression of the ESR signal out of anticrossing, observed
experimentally (Fig.~\ref{fig:5}a). At anticrossing the spin-valley
mixing and the AC driving lift the blockade, making the observation of
ESR possible.

In Fig.~\ref{fig:5}c we plotted the energy difference for each pair of
states, with an intensity given by the absolute value of the AC
coupling which qualitatively reconstructs the anticrossing picture
observed experimentally. Despite the many different transitions which
give rise to multiple ESR lines/crossings, the AC coupling filters out
many of them (still involving all three anticrossing spots, scf.
Fig.~\ref{fig:5}b), that leaves us with only one anticrossing and a
straight line in between, Fig.~\ref{fig:5}c. Actually, just this
picture requires to have the inequality $|a_R| \gg |b_L|$, mentioned
above.

\subsection*{Spin-valley blockade}
\label{Sec:spin_blockade}

It is worth now to consider the observed blockade in the absence of AC
driving and for a finite magnetic field, where the leakage current is
suppressed for the whole region of interdot detuning, $0 \leq
\varepsilon \leq \Delta_{\rm ST}^{\rm exp} \simeq 343\, \mu{\rm eV}$
(see Fig.~\ref{fig:2}c, \ref{fig:2}d), including that which is 2-3
times larger than the valley splitting $E_V \simeq 86.2\, \mu{\rm
  eV}$. This means that a type of spin-valley blockade is
experimentally observed. Since the blockade is magnetic
field-independent for $B \gtrsim 0.5\, {\rm T}$, Fig.~\ref{fig:2}d,
and particularly for $B \sim B_V$, one needs to consider several
possible scenarios of spin-valley blockade. In all scenarios the
blockade means impossibility for an inelastic transition (via phonons)
to happen. For the usual spin-blockade (1) it is since phonon emission
cannot flip spins, and so a triplet $\rm T(1,1)$ cannot decay to the
localized singlet $\rm S(2,0)$. The different type of spin-valley
  blockade, if it happened, is since phonon emission cannot change
the valley content of the state\cite{PalyiBurkard2009PRB} (2), or it
cannot change the ``valley parity'' of the state (3) (related to
specific cancelation of phonon decay amplitudes, see below and
Supplementary Note 3).

\begin{figure}[tbp]
  \centering{\includegraphics[]{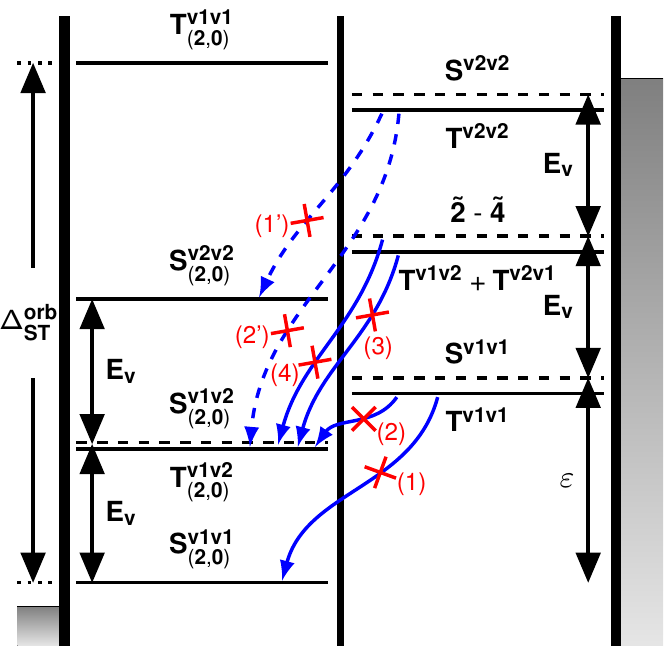}}
  \caption{\textbf{Spin-blockade vs. spin-valley blockade.} Schematics
    of spin-blockade and spin-valley blockade in the case of
    $\Delta_{ST}^{\rm orb} > 2 E_V$, and for interdot detuning
    $\varepsilon > E_V$ (see text). For the energy levels of the
    localized $(2,0)$-states (shown on the left), the states $\rm
    T^{\rm v1v2}_{(2,0)}$ and $\rm S^{\rm v1v2}_{(2,0)}$ are
    degenerate; the `true' singlet-triplet energy splitting
    $\Delta_{ST}^{\rm orb}$ is to the higher orbital (triplet) state
    $\rm T^{\rm v1v1}_{(2,0)}$. The energy levels of the delocalized
    $(1,1)$-states (shown on the right) are fine split by the exchange
    coupling $J$ (see text and Supplementary Note 3). The polarized
    (Zeeman split) $\pm$-triplets of the $(1,1)$-, $(2,0)$-states are
    not shown, and for the $\rm [v_1,v_2]$ group of $(1,1)$-states
    only the relevant blocked states are shown: namely, the two
    polarized triplets $\rm |\tilde{4}_{\pm}\rangle = T_{\pm}^{\rm
      v1v2} + T_{\pm}^{\rm v2v1}$ (shown schematically as one level)
    are referred to the spin-valley blockade (3); the unpolarized
    state $\rm |\tilde{2}\rangle - |\tilde{4}\rangle \equiv (S^{\rm
      v1v2} - S^{\rm v2v1}) - (T_{0}^{\rm v1v2} + T_{0}^{\rm v2v1})$
    is referred to the spin-valley blockade (4) (see also
    Fig.~\ref{fig:5}b). The spin-blockade (1), and the three types of
    possible spin-valley blockade: (2), (3), (4), are explained in the
    text. The blockade from the upper
    valley $(1,1)$-state $\rm T^{\rm v2v2}$ shown as (1') and (2') is
    similar to (1), (2) (see Supplementary Note 3). }
    \label{fig:6}
\end{figure}

It is essential for our argument that all spin-valley states,
Fig.~\ref{fig:6}, are loaded continuously, for any fixed magnetic
field $B$. Assuming the dominance of a phonon inelastic relaxation
(like in a single $\mathrm{Si/SiO_2}$ quantum dot \cite{Yang2013NC}),
we introduce the phonon decay amplitudes between various spin-valley
states\cite{YuCardonaBook}, $a_{\rm LR}^{(ij)} \equiv \langle
\tilde{L}_{\rm v_i}({\bold r})| \hat{H}_{\rm e-ph} |\tilde{R}_{\rm
  v_j}({\bold r}) \rangle,\, i,j=1,2$ (see Supplementary Note 3), and
consider the scenarios: (1) {\it The spin blockade.} For $\varepsilon
< E_V$ and $B \lessgtr B_V$ the polarized triplet states $\rm
T_{\pm,0}^{\rm v1v1}$, are spin-blocked, similar to a GaAs system
\cite{Petta2005S,Koppens2006N} (analogously, the higher valley states
$\rm T_{\pm,0}^{\rm v2v2}$ are spin-blocked as well,
Fig.~\ref{fig:6}). In the region of $B\sim B_V$, the states $\rm
T_{-}^{\rm v1v1}$, $\rm T_{+}^{\rm v2v2}$ still remain blocked. (2)
{\it Spin-valley blockade I.} For larger detuning, $\varepsilon >
E_V$, the spin blockade remains, Fig.~\ref{fig:6}, however the state
$\rm T^{\rm v1v1}$ may decay to the triplet 2-electron state $\rm
T_{(2,0)}^{\rm v1v2}$ in the left dot, since it is energetically
allowed; the blockade will depend on the off-diagonal in valley phonon
decay amplitude $a_{\rm LR}^{(21)}$: for the ideal case $a_{\rm
  LR}^{(21)}=0$ (since umklapp transitions are suppressed, while
  the envelope functions of the states $\rm |v_1\rangle$, $\rm
  |v_2\rangle$ are identical for ideal interface, see Supplementary
  Note 1). The m.e. $a_{\rm LR}^{(21)}$  could be non-zero for
a non-ideal interface, similar to the m.e.
Eqs.~(\ref{dipoleMEa}),~(\ref{dipoleMEb}). Even if this type of valley
blockade is lifted (for $a_{\rm LR}^{(21)}\neq 0$), there could be a
second type of spin-valley blockade, Fig.~\ref{fig:6}. (3) {\it
  Spin-valley blockade II.} The overall blockade could not yet be
lifted, since the spin-valley triplet state $\rm T_0^{\rm v1v2} +
T_0^{\rm v2v1}$ (as well as its polarized counterparts!) cannot decay
to the corresponding $\rm T^{\rm v1v2}_{(2,0)}$ state, which we call a
{\it spin-valley phonon selection rule}. Indeed, the corresponding
diagonal-in-valley phonon decay amplitudes cancel if the equality
holds: $a_{\rm LR}^{(22)} = a_{\rm LR}^{(11)}$. The equality is exact
for an ideal interface, (for identical $\rm v_1, v_2$ envelope
  functions) and is likely to hold at least approximately even for a
non-ideal interface, see Supplementary Note 2. (4) {\it
  Spin-valley blockade III.} Since the blockade is observed at $B \sim
B_V$, one more candidate for a blocked state is the eigenstate (as an
alternative of the blocked state $\rm T_{-}^{\rm v1v1}$) $\rm
|\tilde{2}\rangle - |\tilde{4}\rangle \equiv (S^{\rm v1v2} - S^{\rm
  v2v1}) - (T^{\rm v1v2}_{0} + T^{\rm v2v1}_{0})$, Fig.~\ref{fig:5}b,
Fig.~\ref{fig:6}. In order to be blocked, this requires the above
equality, $a_{\rm LR}^{(22)} = a_{\rm LR}^{(11)}$, and also $a_{\rm
  LR}^{(12)} =0$. As to the blockade alternatives (2-4),
considered here, we notice, that if the $\rm v_1,v_2$ were the lowest
orbital states in each dot, neither of the above equalities would
hold, and the blockade would be lifted just at $\varepsilon > E_V$.

The spin-valley blockade may be lifted for the above alternatives
either at detuning $\varepsilon > E_V$ and/or at $\varepsilon \geq
\Delta_{\rm ST}^{\rm orb} - E_V$. The later is possible when the state
$\rm T^{\rm v1v2} + T^{\rm v2v1}$ matches in energy the usual orbital
state in the left dot, $\rm T_{(2,0)}^{\rm orb,v1v1}$ (see
Fig.~\ref{fig:6}, Supplementary Figure 4 and Supplementary Note 4).
Since in the experiment the inequality holds: $\Delta_{\rm ST}^{\rm
  orb} > 2 E_V \simeq 172\, \mu{\rm eV}$, the states match at an
energy larger than $E_V$, namely at the experimentally observed
$\varepsilon = \Delta_{\rm ST}^{\rm exp}= 343\, \mu{\rm eV}$, which
implies a true S-T splitting
\begin{equation}
\Delta_{\rm ST}^{\rm orb} = \Delta_{\rm ST}^{\rm exp} +  E_V \simeq 439\, \mu{\rm eV}
\label{Delta_orb} .
\end{equation}
The measured experimental blockade cannot distinguish which of the
above alternatives has happened. However, the single fact that we have
observed a blockade of the leakage current for detunings $\varepsilon
> E_V$ allows us to state that we have observed the spin-valley
  blockade associated with either of the alternatives (2), (3), (4).

\section*{DISCUSSION}
\label{sec:conclusion}

In conclusion, we have observed the electron spin resonance using spin
or spin-valley blockade in a gate defined Si MOS DQD. The ESR signal
is significantly enhanced where Zeeman levels of different valley
content (anti)cross, due to the spin-valley mixing arising from
spin-orbit interaction at non-ideal quantum dot interfaces. From the
ESR linewidth, the spin dephasing time is estimated as $T_{2}^{*}
\simeq 63\, {\rm ns}$, which is significantly longer than in a GaAs
system. The discovery of the anomalous anticrossing demonstrates the
possibility to characterize and manipulate spin-valley states using
ESR, for individual qubits. In a long run, with a
  better understanding of the device physics of silicon quantum dots,
  one can choose, design, and operate qubits in regimes which are
  better suited for robust quantum computation (examples could include
  making valley splitting large enough across devices or improving
  surface interfaces). Our results improve the outlook of Si MOS QDs
as a platform for high-coherence spin qubits, in the now leading
microelectronics material.

\section*{METHODS}

\subsection*{Device fabrication}
\label{sec:device-fabrication}

The sample used in this experiment was fabricated on undoped
commercial $\mathrm{Si}$ wafer with a $50\,{\rm nm}$ thermal
$\mathrm{SiO}_{2}$ \cite{Xiao2010PRL,Xiao2010APL}. First, the ohmic
contacts were made by phosphorous ion-implantation followed by a high
temperature annealing. Then, the confinement gates as well as the
co-planar strip (CPS) loop were defined by electron beam lithography.
Before putting on the global accumulation gate (Cr/Au), a $120\,{\rm
  nm}$ $\mathrm{Al}_{2}\mathrm{O}_{3}$, which serves as an insulating
layer between confinement gates and accumulation gates, was grown by
atomic layer deposition. See Fig.~\ref{fig:1}a for the cross section
layout of the DQD device. We use aluminum as the material for CPS loop
and obtain a few $\Omega$ loop resistance to maximize the transmission
of microwave signal.

\subsection*{Electrical measurement}
\label{sec:measurement}

The device is mounted on the cold-finger of a dilution refrigerator
with a base temperature of $80\,{\rm mK}$. An in-plane magnetic field
is created via a superconducting magnet; possible trapping of a
residual magnetic flux may cause an overall shift of the magnetic
field read-off by a few mT. The electron temperature is about
$200\,{\rm mK}$. A semirigid cable delivers the microwave, which is
generated by HP/Agilent Signal Generator 8673B, to the coplanar loop.
The cable has an attenuation of $20\,{\rm dB}$ at the frequency of
$20\, {\rm GHz}$. A low-noise current amplifier ($5\,{\rm
  fA}/\sqrt{{\rm Hz}}$) is used to measure the quantum dot transport
current.

\section*{ACKNOWLEDGMENTS}       
The authors thank Xuedong Hu, Bill Coish, Dimitrie Culcer, Andrew
Dzurak, Andrew Hunter, Sue Coppersmith, John Gamble, Mark Friesen, and
Andrey Kiselev for helpful discussions. The work is sponsored by the
U.S. Department of Defense and by the U. S. Army Research Office
(w911NF-11-1-0028).

\section*{FINANCIAL INTERESTS}
The authors declare no competing financial interests.

\section*{AUTHOR CONTRIBUTIONS}
X.H. performed the experiment and analyzed the data. M.X. designed and
fabricated the sample. R.R. and C.T. developed the theory of
spin-valley mixing. H.W.J. supervised the project. All authors
contributed to the data interpretation and to the paper writing.

\end{document}